\documentclass[a4paper,3p]{elsarticle}
\usepackage[utf8]{inputenc}
\usepackage{amsmath}
\usepackage{amsfonts}
\usepackage{graphicx}
\usepackage{float}
\usepackage{bm}
\usepackage{hyperref}
\usepackage{color}
\usepackage{amssymb}
\usepackage{dsfont}
\usepackage{slashed}
\usepackage{mathtools}
\usepackage[compat=1.1.0]{tikz-feynman}
\usetikzlibrary{positioning}
\tikzset{every picture/.style={scale=0.6, transform shape}}
\tikzset{node distance=0.7cm and 0.7cm}  
\tikzfeynmanset{
  arrow size=1pt,
  every edge={line width=0.5pt}
}

\usepackage{hyperref}
\hypersetup{
colorlinks=true,
urlcolor=blue,
linkcolor = blue,
urlcolor  = blue,
citecolor = blue,
anchorcolor = blue,
pdftitle={$g_A$ at two loops},
pdfsubject={$g_A$ at two loops}}
\allowdisplaybreaks
\makeatletter
\def\ps@pprintTitle{%
\let\@oddhead\@empty
\let\@evenhead\@empty
\def\@oddfoot{}%
\let\@evenfoot\@oddfoot}
\makeatother

\begin{document}
\title{Calculation of  the axial-vector coupling constant $g_A$ to two loops\\ in covariant chiral perturbation theory}
\author[orsay]{V\'eronique Bernard}
\author[rub,tsu]{Jambul Gegelia}
\author[bonn]{Shayan Ghosh}
\author[bonn,fzj,beihang]{Ulf-G. Mei\ss{}ner}
\address[orsay]{Universit\'e Paris-Saclay, CNRS/IN2P3, IJCLab, 91405 Orsay, France}
\address[rub]{Ruhr-Universit\"at Bochum, Fakult\"at f\"ur Physik und Astronomie, Institut f\"ur Theoretische Physik II, D-44780 Bochum, Germany}
\address[tsu]{Tbilisi State University, 0186 Tbilisi, Georgia}
\address[bonn]{Helmholtz-Institut f\"ur Strahlen- und
Kernphysik and Bethe Center for Theoretical Physics,\\
Universit\"at Bonn, D-53115 Bonn, Germany}
\address[fzj]{Institute for Advanced Simulation (IAS-4), Forschungszentrum J\"ulich,
D-52425 J\"ulich, Germany}
\address[beihang]{Peng Huanwu Collaborative Center for Reserch and Education, International Institute for Interdisciplinary and Frontiers, Beihang University, Beijing 100191, China}
\begin{abstract}
We present a  calculation of the leading two-loop corrections to the axial-vector
coupling constant $g_A$ in two covariant versions of two-flavor baryon chiral perturbation theory.
Taking the low-energy constants from a combined analysis of elastic and  inelastic pion-nucleon
scattering, we find that these corrections are rather moderate. 
\end{abstract}
\maketitle

\section{Introduction}
\label{sec:intro}

 The nucleon axial-vector coupling constant $g_A$  plays
an eminent role in neutron $\beta$-decay, see Ref.~\cite{Czarnecki:2018okw} for a recent work,
in nuclear reactions and in neutrino scattering off nuclei, as witnessed by the  T2K, NOvA, MINERvA,
MicroBooNE, and SBN experiments.
For reviews, see Refs.~\cite{Bernard:2001rs,Suhonen:2017krv}.
Furthermore, $g_A$ is also a fundamental parameter of  low-energy
chiral QCD dynamics which parameterizes the leading order pion-nucleon interaction as at
tree level the Goldberger-Treiman relation (GTR) allows to express the pion-nucleon coupling
in terms of the axial-vector coupling. Corrections to the GTR are known to be small so that this
relation is an approximate one when considering loops.
  Furthermore, the axial-vector coupling  is also
considered a ``gold-plated'' observable for the {\em ab initio} lattice QCD approach. The first
high-precision lattice QCD calculation was reported in Ref.~\cite{Chang:2018uxx} and a summary of older and more 
recent results is collected in Ref.~\cite{FlavourLatticeAveragingGroupFLAG:2024oxs}. Most of these
simulations are performed for unphysical pion masses, while in
Refs.~\cite{Alexandrou:2024ozj,QCDSFUKQCDCSSM:2023qlx,Jang:2023zts,Hall:2025ytt}
the physical pion mass is also considered. In any case,
the issue of precise and controlled chiral extrapolations is still of relevance.
Of course, it is also of general interest to study the higher-order corrections in the
quark mass expansion of this observable,
as they encode information about the convergence of two-flavor baryon chiral perturbation theory.
Therefore, in this work, we  consider the calculation of $g_A$ to two loops using two different 
covariant schemes, namely the extended-on-mass-shell (EOMS) approach~\cite{Fuchs:2003qc} and 
the method of infrared regularization (IR)~\cite{Becher:1999he}.
Detailed discussions of these schemes and comparisons with the often used heavy baryon approach can 
be found in Refs.~\cite{Bernard:2007zu,Meissner:2022cbi}. This work differs from the earlier paper~\cite{Bernard:2006te},
where renormalization group methods were applied to the chiral pion-nucleon Lagrangian in the heavy
baryon approach and the two-loop representation was confronted with then existing lattice calculations
at unphysical pion masses. We will come back to that work later. 

This article is organized as follows. In Sec.~\ref{sec:exp} we present the
general form of the chiral expansion of $g_A$. Sect.~\ref{sec:form} outlines
the calculations of the two-loop corrections. We present and discuss our
results in Sec.~\ref{sec:res}. \ref{app:A} contains some further results.

\section{Chiral expansion of $g_A$}
\label{sec:exp}

In baryon chiral perturbation theory (BCHPT), one encounters odd and even powers of the small expansion
parameter $q$, which in the two flavor case is given by pion masses and momenta as well as nucleon
three-momenta.
The nucleon mass $m_N$ is of the same size as the hard scale related to chiral symmetry breaking, often
estimated as $\Lambda_\chi = 4\pi F_\pi \simeq 1.2\,$GeV, with $F_\pi \simeq 92\,$MeV the pion decay
constant.  Tree diagrams start contributing 
at order $q$, one-loop diagrams at order $q^3$,  two-loop ones at order $q^5$, three-loop diagrams
at order $q^7$, and so on. In this paper we will concentrate on the leading two-loop diagrams of
chiral order $q^5$, the  order $q^6$ two-loop calculation is in progress~\cite{bggmlong}. The
latter involves the calculation of 
several two-loop diagrams with vertices from ${\cal L}_{\pi N}^{(2)}$.
Consequently, the chiral expansion of the axial-vector coupling 
as evaluated in this work takes the form
\begin{eqnarray}
\label{gAstruc}
g_A &=& g_0 \,\, \biggl\{ 1 + \left( \frac{\alpha_2}{(4\pi F)^2} \ln
\frac{M}{\mu} + \beta_2 \right) \, M^2 + \alpha_3 \, M^3
\nonumber\\
&& \quad +  \left(\frac{\alpha_4}{(4\pi F)^4} \ln^2\frac{M}{\mu}
+  \frac{\gamma_4}{(4\pi F)^4} \ln\frac{M}{\mu}  + \beta_4
\right) \, M^4 \biggr\} + {\cal O}(M^5) ~,
\nonumber \\
&=&  g_0 \,\, \biggl\{ 1 + \underbrace{\Delta^{(2)} + \Delta^{(3)}}_{\rm  one-loop}
  + \underbrace{\Delta^{(4)}}_{\rm leading~two-loop} 
    \biggr\}   + {\cal O}(M^5)~,
    \label{eq:gaexp}
\end{eqnarray}
with $g_0$ the chiral limit value of $g_A$, 
\begin{equation}
g_A = g_0\, \left[ 1 + {\cal O}(M^2)\right]\, ,
\end{equation}
and $\Delta^{(4)}$ contains the two loop contribution of chiral order $q^5$, as well as contributions from 
one loop diagrams with vertices from ${\cal L}_{\pi N}^{(1,2,3)}$. Note that terms with denominations one-loop/leading two-loop in the last line of 
Eq.~\eqref{eq:gaexp} include the contributions from tree diagrams with  vertices from ${\cal L}_{\pi N}^{(3/5)}$.
Further, $\mu$ is the scale of dimensional regularization and $\Delta^{(n)}$ represents the correction at order $n$ in the chiral counting.
We have expressed  $g_A$ in terms of the pion decay constant in the chiral limit, $F$,  and the leading order
term of the quark mass expansion of the pion mass squared, $M^2$. One has:
\begin{equation}
F_\pi= F\,\left[1 +{\cal O}(M^2)\right] \, ,  \quad M_\pi^2 = M_{}^2\, \left[1 +{\cal O}(M^2)\right]~.
\end{equation}
The difference between $F$ and $F_\pi$  ($M$ and $M_\pi$) is of higher order when working at ${\cal O}(q^3)$, 
however, at the order we are working it has to be taken into  account. 
The one-loop coefficients 
$\alpha_2$ and $\beta_2$ in Eq.~(\ref{gAstruc}) were first given 
in Ref.~\cite{Bernard:1992qa} and the one-loop fourth order calculation 
was completed in Ref.~\cite{Kambor:1998pi} with  (see also Ref.~\cite{Schindler:2006it}),
\begin{eqnarray}\label{oneloop}
\alpha_2 &=& -2 -4g_0^2~, \nonumber\\ 
\beta_2 &=& \frac{4}{g_0} {d}_{16}^r(\mu) 
 - \frac{g_0^2}{(4\pi F)^2}~, 
\nonumber\\ 
\alpha_3 &=& \frac{1}{24\pi F^2 m}
\left(3+3g_0^2-4m c_3+8m c_4\right)~,
\label{eq:co23}
\end{eqnarray}
with $m$ the nucleon mass in the two-flavor chiral limit (sometimes also denoted as  $m_0$).  Further,
the expressions in Eq.~\eqref{eq:co23}  are identical in the IR and EOMS schemes when expanded in powers of  $M$.
Indeed in the latter
the additional so-called regular terms, which violate the power counting, can be absorbed into the LECs
$g_0$ and $d_{16}$, see Ref.~\cite{Siemens:2016hdi}  and below. One has in the EOMS scheme
\begin{equation}
g_0 {\color{blue} \to } \, g_0 + \delta g_0|_{\rm reg} \,  \quad d_{16} {\color{blue} \to } \, d_{16} + \delta d_{16}|_{\rm reg}~,
\end{equation}
where the subscript ``reg'' denotes the  above mentioned regular terms.

 Let us now discuss the $M^4$ contribution to $g_A$.
Note that contrary to the work~\cite{Bernard:2006te}, in which {\color{blue} $\alpha_4$}
was calculated in heavy baryon chiral perturbation theory using renormalization group methods, we use
here the operator basis and low-energy constants (LECs) given in Ref.~\cite{Gasser:2002am}.
There are four types of contributions to $\Delta^{(4)}$: the  irreducible diagrams,
the one coming from the
wave function  and coupling constant renormalization, the one loop graphs with  vertices from
${\cal L}_{\pi N}^{(3)}$, and counterterms. One thus has schematically 
\begin{equation}
  g_0 \, \Delta^{(4)} 
 =  g_A^{\rm ren} + g_A^{d_i} + g_A^{\rm irr}  +g_A^{\rm ct}~.
 \label{eq:ga2loop}
\end{equation}

\section{Outline of the calculation}
\label{sec:form}

Here, we sketch the calculation of the four different contributions mentioned above. For all the details,
we refer to Ref.~\cite{bggmlong}.

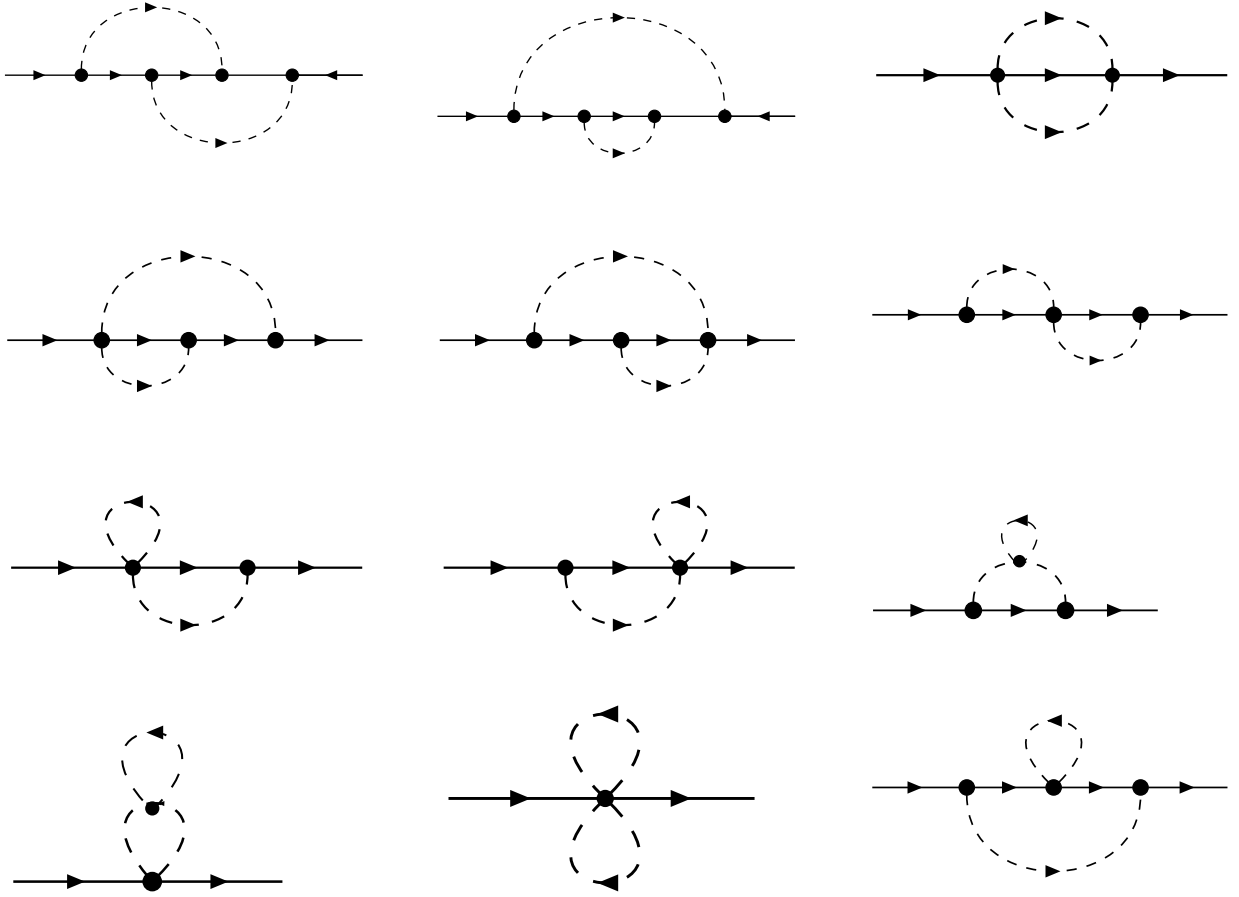
\begin{figure}[htb]
\begin{minipage}{0.32\textwidth}
 \resizebox{0.925\textwidth}{!}{
\begin{tikzpicture}
\begin{feynman}
	\vertex(a) ;
	\node[right=of a, dot, scale=1.75] (b) {};
	\node[right=of b, dot, scale =1.75] (c) {};
	\node[right=of c, dot, scale=1.75] (d) {};
	\node[right=of d, dot, scale=1.75] (e) {};
	\vertex[right=of e] (f) ;
	
\diagram* {
	(a)-- [fermion, arrow size=1pt] (b),
	(b) -- [fermion, arrow size=1pt] (c),
	(b) -- [charged scalar, half left, arrow size=1pt] (d),
	(c) -- [charged scalar, half right, arrow size=1pt] (e),
	(c) -- [fermion, arrow size=1pt] (d),
	(d)-- [fermion, arrow size=1pt] (f),
	(f)-- [fermion, arrow size=1pt] (e)  };
\end{feynman}
\end{tikzpicture}}
\end{minipage}
\hspace{0.3truecm}
\begin{minipage}{0.32\textwidth}
 \resizebox{0.925\textwidth}{!}{
\begin{tikzpicture}
\begin{feynman}
	\vertex(a) ;
	\node[right=of a, dot, scale=1.75] (b) {};
	\node[right=of b, dot, scale =1.75] (c) {};
	\node[right=of c, dot, scale=1.75] (d) {};
	\node[right=of d, dot, scale=1.75] (e) {};
	\vertex[right=of e] (f) ;
	
\diagram* {
	(a)-- [fermion, arrow size=1pt] (b),
	(b) -- [fermion, arrow size=1pt] (c),
	(b) -- [charged scalar, half left, arrow size=1pt] (e),
	(c) -- [charged scalar, half right, arrow size=1pt] (d),
	(c) -- [fermion, arrow size=1pt] (d),
	(d)-- [fermion, arrow size=1pt] (f),
	(f)-- [fermion, arrow size=1pt] (e)  };
\end{feynman}
\end{tikzpicture}}
\end{minipage}
\hspace{0.3truecm}
\begin{minipage}{0.32\textwidth}
 \resizebox{0.925\textwidth}{!}{
\begin{tikzpicture}
\begin{feynman}
	\vertex(a) ;
	\node[right=of a, dot, scale=1.15] (b) {};
	\node[right=of b, dot, scale=1.15] (c) {};
	\vertex[right=of c](d) ;
	
\diagram* {
	(a)-- [fermion, arrow size=0.85pt] (b),
	(b) -- [fermion, arrow size=0.85pt] (c),
	(b) -- [charged scalar, half left, arrow size=0.85pt] (c),
	(b) -- [charged scalar, half right, arrow size=0.85pt] (c),
	(c) -- [fermion, arrow size=0.85pt] (d) };
\end{feynman}
\end{tikzpicture}}
\end{minipage}
\vspace{0.8truecm}

\begin{minipage}{0.32\textwidth}
 \resizebox{0.925\textwidth}{!}{
\begin{tikzpicture}
\begin{feynman}
	\vertex(a) ;
	\node[right=of a, dot, scale=1.75] (b) {};
	\node[right=of b, dot, scale =1.75] (c) {};
	\node[right=of c, dot, scale =1.75] (d) {};
	\vertex[right=of d] (e) ;
	
\diagram* {
	(a)-- [fermion, arrow size=1pt] (b),
	(b) -- [fermion, arrow size=1pt] (c),
	(b) -- [charged scalar, half left, arrow size=1pt] (d),
	(b) -- [charged scalar, half right, arrow size=1pt] (c),
	(c) -- [fermion, arrow size=1pt] (d),
	 (d) -- [fermion, arrow size=1pt] (e)};
\end{feynman}
\end{tikzpicture}}
\end{minipage}
\hspace{0.3truecm}
\begin{minipage}{0.32\textwidth}
 \resizebox{0.925\textwidth}{!}{
\begin{tikzpicture}
\begin{feynman}
	\vertex(a) ;
	\node[right=of a, dot, scale=1.75] (b) {};
	\node[right=of b, dot, scale =1.75] (c) {};
	\node[right=of c, dot, scale =1.75] (d) {};
	\vertex[right=of d] (e) ;
	
\diagram* {
	(a)-- [fermion, arrow size=1pt] (b),
	(b) -- [fermion, arrow size=1pt] (c),
	(b) -- [charged scalar, half left, arrow size=1pt] (d),
	(c) -- [charged scalar, half right, arrow size=1pt] (d),
	(c) -- [fermion, arrow size=1pt] (d) ,
	(d) -- [fermion, arrow size=1pt] (e)  };
\end{feynman}
\end{tikzpicture}}
\end{minipage}
\hspace{0.3truecm}
\begin{minipage}{0.32\textwidth}
 \resizebox{0.925\textwidth}{!}{
\begin{tikzpicture}
\begin{feynman}
	\vertex(a) ;
	\node[right=of a, dot, scale=1.75] (b) {};
	\node[right=of b, dot, scale =1.75] (c) {};
	\node[right=of c, dot, scale =1.75] (d) {};
	\vertex[right=of d] (e) ;
	
\diagram* {
	(a)-- [fermion, arrow size=0.9pt] (b),
	(b) -- [fermion, arrow size=0.9pt] (c),
	(c) -- [fermion, arrow size=0.9pt] (d),
	(b) -- [charged scalar, half left, arrow size=0.8pt] (c),
	(c) -- [charged scalar, half right, arrow size=0.8pt] (d),
	(d) -- [fermion, arrow size=0.9pt] (e) };
\end{feynman}
\end{tikzpicture}}
\end{minipage}
\vspace{1.truecm}

\begin{minipage}{0.32\textwidth}
 \resizebox{0.925\textwidth}{!}{
\begin{tikzpicture}
\begin{feynman}
\vertex(a) ;
	\node[right=of a, dot, scale=1.25] (b) {};
	\node[right=of b, dot, scale=1.25] (c) {} ;
	\vertex[right=of c] (d) ;
	
\diagram* {
	(a) -- [fermion, arrow size=0.9pt] (b),
	(b) -- [fermion, arrow size=0.9pt] (c),
	b -- [charged scalar, out=45, in=135, min distance=1.5cm, arrow size=0.8pt] b,
	b -- [charged scalar, half right, arrow size=0.8pt]c,
	(c)-- [fermion,arrow size=0.9pt] (d) };	
\end{feynman}
\end{tikzpicture}}
\end{minipage}
\hspace{0.3truecm}
\begin{minipage}{0.32\textwidth}
 \resizebox{0.925\textwidth}{!}{
\begin{tikzpicture}
\begin{feynman}
\vertex(a) ;
	\node[right=of a, dot, scale=1.25] (b) {};
	\node[right=of b, dot, scale =1.25] (c) {};
	\vertex[right=of c] (d);
	
\diagram* {
	(a)-- [fermion, arrow size=0.9pt] (b),
	(b) -- [fermion, arrow size=0.9pt] (c),
	c -- [charged scalar, out=45, in=135, min distance=1.5cm, arrow size=0.8pt] c,
	b -- [charged scalar, half right, arrow size=0.8pt]c,
	(c) -- [fermion, arrow size=0.9pt] (d)};
	
\end{feynman}
\end{tikzpicture}}
\end{minipage}
\hspace{0.3truecm}
\begin{minipage}{0.32\textwidth}
 \resizebox{0.75\textwidth}{!}{
\begin{tikzpicture}
\begin{feynman}
	\vertex(a) ;
	\node[right=of a, dot, scale=1.75] (b) {};
	\node[right=of b, dot, scale =1.75] (c) {};
	\vertex[right= of c ] (d); 
	\vertex at ($(b)!0.5!(c)+(0,0.8)$) (bc);

\diagram* {
	(a)-- [fermion, arrow size=1pt] (b),
	(b) -- [fermion, arrow size=1pt] (c),
	(c) -- [fermion, arrow size=1pt] (d),
	(b) -- [charged scalar, half left, arrow size=0pt] (c),			
 };
 \path (b) .. controls +(0.86, 0.98) and +(-0.86, 0.86) .. (c) coordinate[pos=0.5] (mid); 
\draw[charged scalar, half left, arrow size=0.9pt]  (mid) .. controls +(1., 1.) and +(-1., 1.) .. (mid);
 \filldraw[fill=black ] (bc) circle [radius=0.09cm]; 
\end{feynman}
\end{tikzpicture}}
\end{minipage}
\vspace{0.5truecm}

\begin{minipage}{0.32\textwidth}
 \resizebox{0.725\textwidth}{!}{
\begin{tikzpicture}
\begin{feynman}
	\vertex(a) ;
	\node[right=of a, dot, scale=1.35] (b) {};	
	\vertex[right= of b ] (c); 
	\node[above= of b] (d);

\diagram* {
	(a)-- [fermion, arrow size=0.8pt] (b),
	(b) -- [fermion, arrow size=0.8pt] (c),
	b -- [charged scalar, out=45, in=130, min distance=1.5cm, arrow size=0.2pt] b	
 };
 \path (b) .. controls +(0.95, 1.18) and +(-0.95, 0.95) .. (b) coordinate[pos=0.5] (mid); 
\draw[charged scalar,  out=45, in=130, min distance=1.5cm, arrow size=0.75pt
]  (mid) .. controls +(1.2, 1.2) and +(-1.2, 1.2) .. (mid);
\path (b) .. controls +(1.1, 1.1) and +(-1.1, 1.1) .. (b) coordinate[pos=0.5] (mid1);
 \filldraw[fill=black ] (mid1) circle [radius=0.07cm]; 
\end{feynman}
\end{tikzpicture}}
\end{minipage}
\hspace{0.3truecm}
\begin{minipage}{0.32\textwidth}
 \resizebox{0.825\textwidth}{!}{
\begin{tikzpicture}
\begin{feynman}
\vertex(a) ;
	\node[right=of a, dot, scale=1.] (b) {};
	\vertex[right=of b] (c) ;
	\vertex[above right=of b] (g) ;
	
\diagram* {
	(a) -- [fermion, arrow size=0.8pt] (b),
	(b) -- [fermion, arrow size=0.8pt] (c),
	b -- [charged scalar, out=45, in=135, min distance=1.5cm, arrow size=0.8pt] b,
	b -- [charged scalar, out=-45, in=-135, min distance=1.5cm, arrow size=0.8pt] b
	 };	
\end{feynman}
\end{tikzpicture}}
\end{minipage}
\hspace{0.3truecm}
\begin{minipage}{0.32\textwidth}
 \resizebox{0.925\textwidth}{!}{
\begin{tikzpicture}
\begin{feynman}
\vertex(a) ;
	\node[right=of a, dot, scale=1.75] (b) {};
	\node[right=of b, dot, scale =1.75] (c) {};
	\node[right=of c, dot, scale=1.75] (d) {};
	\vertex[right=of d] (e);
	\vertex[below=of c] (f) {  };
	
\diagram* {
	(a)-- [fermion, arrow size=1pt] (b),
	(b) -- [fermion, arrow size=1pt] (c),
	(c) -- [fermion, arrow size=1pt] (d),
	(d) -- [fermion, arrow size=1pt] (e),
	c -- [charged scalar, out=45, in=135, min distance=2cm, arrow size=1pt] c,
	(b) -- [charged scalar, half right, arrow size=1pt] (d)
};
\end{feynman}
\end{tikzpicture}}
\end{minipage}
\caption{Two-loop diagrams contributing to the nucleon self-energy. The same topologies contribute
  also to $g_A$.
  For the latter one has to hook  the axial current  wherever possible on the nucleon propagator and on the
  vertices with pions and nucleons.  The dashed lines represent the pion and the  solid ones the nucleon.
  The dots are vertices from ${\cal L}_{\pi N}^{(1)}$  at the order we are working.
  Note that there are further two-loop diagrams shown in Fig.~\ref{fig2}
    where the axial-current couples to three pions.
  }
\label{fig1}
\end{figure}

Let us first discuss the wave function and coupling constant renormalization. The  wave function
renormalization constant $Z$ is the residue of the pole in the two point function and is determined by
\begin{equation}
   Z^{-1} =1-\frac {d}{d {\slashed  p} }\Sigma(\slashed  p)\bigg|_{\slashed p=m_N} 
 \end{equation}
 with $\Sigma(\slashed p)$ the nucleon self-energy, which has a similar expansion as discussed above for $g_A$, namely
 \begin{equation}
Z=1 + Z_{\rm 1-loop} +Z_{\rm 2-loop} + \cdots \,,
\end{equation}
where one has 
\begin{eqnarray}\label{eq:Z1l}
 Z_{\rm 1-loop}&=&-\frac{9}{32  \pi^2 F^2} M^2  g_0^2\biggl \{ 16 \pi^2 \lambda +  \log  \frac{M}{\mu}+\frac{1}{ 3}-\frac{ \pi M}{2 m}
 -\epsilon \biggl[\log^2  \frac{M}{\mu} +\frac{2}{ 3} \log  \frac{M}{\mu} \nonumber\\
 && \qquad\qquad\qquad\qquad \qquad\qquad
 +\frac{1}{24}(6 +\pi^2)- \frac{\pi}{m}M \biggl(\log  \frac{M}{\mu} -\frac{1}{6}\left(1+6\log 2 \right)\biggr)\biggr] \biggr\}
 \nonumber \\
 &&+\frac{3}{16  \pi^2 F^2} M^2  g_0^2\biggl\{ 16 \pi^2 \lambda +  \log  \frac{m}{\mu}-\frac{1}{ 2}-\epsilon \biggl[ \log^2  \frac{m}{\mu} - \log  \frac{m}{\mu} +\frac{1}{4}\biggl(5 +\frac{\pi^2}{6}\biggr)\biggr]\biggr \}
 \end{eqnarray}
 with
 \begin{equation}
\lambda=\frac{1}{16 \pi^2}\biggl(\frac{1}{d-4}-\frac{1}{2}\bigl(\log (4 \pi)-\gamma_E +1 \bigr)\biggr) \equiv -\frac{1}{32 \pi^2} \lambda_0~,
\end{equation}
and  $\gamma_E$ is the Euler-Mascheroni constant. The first two lines correspond to the pure infrared result
while the last line gives the contribution from the regular part. As expected the latter exhibits only
analytic terms in $M^2$.  The total sum given in Eq.~\eqref{eq:Z1l} is the EOMS result expanded up to $M^3$.
Note that one needs
the results of the one-loop calculation up to 
order ${\cal O}(\epsilon)$  as it contributes to the product of two one-loop quantities. Further, $d=4-2\epsilon$
is the dimension of the space-time.
$Z_{\rm 2-loop}$ can be obtained from the nucleon self-energy  to two loops, whose
purely IR part is given in Ref.~\cite{Schindler:2007dr} and  in
Refs.~\cite{Conrad:2024sla,Conrad:2024phd,Chen:2024twu,Liang:2025cjd} within the EOMS scheme. In order to
calculate the $Z$ factor we have performed the calculation of $\Sigma(p^2)$  at  two loop order, see
the diagrams in Fig.~\ref{fig1}, and found  agreement with these works at the order we are working.
We thus finally have
\begin{equation}
g_A^{\rm ren}=g_0\,\left(\left(\Delta^{(2)} + \Delta^{(3)}\right) Z_{\rm 1-loop}
+ g_0 \left( Z_{\rm 1-loop}^2 +Z_{\rm 2-loop}\right)\right)~.
\end{equation}

\begin{figure}[t]
\begin{minipage}{0.25\textwidth}
\begin{tikzpicture}
\begin{feynman}
	\vertex(a) ;
    \vertex[right= of a](b);
    \vertex[above=of b](d);
    \vertex[above=of d](e);
    \vertex[right= of b] (c);
	\node[right=of a, dot, scale=1.75] (b) {};
	
\diagram* {
     (a) --[fermion] (b),
    (b) --[fermion] (c),
    (b) --[charged scalar] (d),
    (b) --[charged scalar,quarter left] (d),
    (d) --[charged scalar,quarter left] (b),
    (d) --[boson] (e)};
\end{feynman}
\end{tikzpicture}
\end{minipage}
\hspace{-1.2truecm}
\begin{minipage}{0.25\textwidth}
\begin{tikzpicture}
\begin{feynman}
	\vertex(a) ;
    \vertex[right= of a](b);
    \vertex[right= of b] (c);
    \vertex[right= of c] (f);
   \vertex at ($(b)!0.5!(c)$)(g);
   \vertex[above= of g](d);
    \vertex[above=of d](e);   
	\node[right=of a, dot, scale=1.75] (b) {};
	\node[right=of b, dot, scale =1.75] (c) {};
	
\diagram* {
     (a) --[fermion] (b),
    (b) --[fermion] (c),
    (c) --[fermion] (f), 
    (b) --[charged scalar] (d),
    (b) --[charged scalar,quarter left] (d),
    (d) --[charged scalar,quarter left] (c),
    (d) --[boson] (e)};
\end{feynman}
\end{tikzpicture}
\end{minipage}
\hspace{0.2truecm}
\begin{minipage}{0.25\textwidth}
\begin{tikzpicture}
\begin{feynman}
	\vertex(a) ;
    \vertex[right= of a](b);
    \vertex[right= of b] (c);
    \vertex[right= of c] (f);
   \vertex at ($(b)!0.5!(c)$)(g);
   \vertex[above= of g](d);
    \vertex[above=of d](e);   
	\node[right=of a, dot, scale=1.75] (b) {};
	\node[right=of b, dot, scale =1.75] (c) {};
	
\diagram* {
     (a) --[fermion] (b),
    (b) --[fermion] (c),
    (c) --[fermion] (f), 
    (b) --[charged scalar,quarter left] (d),
    (c) --[charged scalar] (d),
    (d) --[charged scalar,quarter left] (c),
    (d) --[boson] (e)};
\end{feynman}
\end{tikzpicture}
\end{minipage}
\hspace{0.2truecm}
\begin{minipage}{0.25\textwidth}
\begin{tikzpicture}
\begin{feynman}
	\vertex(a) ;
    \vertex[right= of a](b);
    \vertex[right= of b] (c);
     \vertex[above=of c](d);
    \vertex[above=of d](e);
    \vertex[right= of c] (f);
    \vertex[right= of f] (g);
	\node[right=of a, dot, scale=1.75] (b) {};
	\node[right=of b, dot, scale =1.75] (c) {};
	\node[right=of c, dot, scale=1.75] (f) {};
	
\diagram* {
     (a) --[fermion] (b),
    (b) --[fermion] (c),
    (c) --[fermion] (f), 
    (f) --[fermion] (g),
    (b) --[charged scalar] (d),
    (c) --[charged scalar] (d),
    (f) --[charged scalar] (d),
    (d) --[boson] (e)};
\end{feynman}
\end{tikzpicture}
\end{minipage}
\vspace{0.5truecm}

\caption{Two-loop diagrams contributing to $g_A$.  The wavy line represents the axial current, the dashed ones
  the pion and the  solid lines the nucleon.}
\label{fig2}
\end{figure}
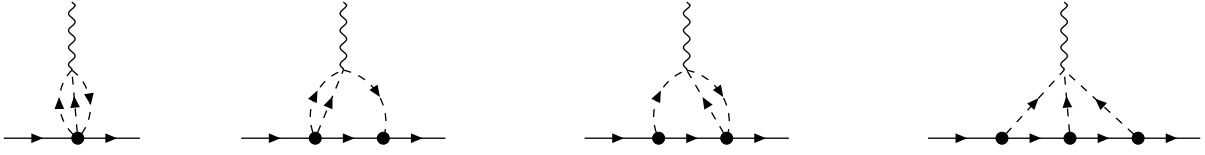

Let us turn to $g_A^{\rm irr}$. 
There are 44  irreducible diagrams contributing at two loop order. Forty
have the same topologies as the ones for the nucleon mass with the axial 
current interacting with the nucleon and up to four pions wherever possible.
For example  the first graph in Fig.~\ref{fig1} leads to 7 diagrams, the second and third
ones to 8 diagrams each, and so on. 
They are proportional to $g_0^{1,3,5}$.  Additionally there are  four  diagrams  specific
to the calculation of the axial current, where the axial current couples to three pions, see Fig.~\ref{fig2}.
In order to calculate these 44 graphs we used the  Mathematica program Feyncalc  \cite{Shtabovenko:2020gxv},
TARCER \cite{Mertig:1998vk} and finally  HypExp to expand  the Hypergeometric functions \cite{Huber:2007dx}.
It allows to write the result in terms of a small set of scalar master integrals  
$F_{\alpha \beta \gamma \delta \epsilon}(m_1,m_2,m_3,m_4,m_5)$ with two integration variables involving up to
five propagators: 
\begin{eqnarray}
&&  F_{\alpha \beta \gamma \delta \epsilon}(m_1,m_2,m_3,m_4,m_5) 
\\  
&& \hspace{-0.8truecm}= \int \frac{d^d k}{(2\pi)^d} \frac{d^dk'}{(2\pi)^d} 	
	\frac{1}
	{ [k^2-m_1^2]^\alpha [k'^2-m_2^2]^\beta[(k-p)^2 - m_3^2]^\gamma[(k'-p)^2 - m_4^2]^\delta[(k-k')^2 - m_5^2]^\epsilon}~,
    \nonumber
\end{eqnarray}
with $m_i$ denoting here either the nucleon or the pion mass. These functions also depend
in principle on $p^2$, but unless specified otherwise they are understood to be taken at $p^2=m^2$ in the following. Note that when $\beta=0$  and $\gamma=0$ these functions are the well-known sunset integrals
which have been rather well studied in the literature, 
using the Mellin-Barnes representation, see for example Ref.~\cite{Ananthanarayan:2020fhl} (and references therein).
In particular the full $\epsilon$-dependent expression of the  functions $ F_{1,0,0,1,1}(M,m,M)$ and
$F_{2,0,0,1,1}(m,M,M)$ to all orders in $M$ is given in Ref.~\cite{ABBFGM} providing  some checks of our results. 
Following Ref.~\cite{Schindler:2007dr} we split these loop functions into a purely infrared part $F_4$,
a mixed term $F_2 +F_3$ and  a regular part $F_1$:
\begin{equation}
  F_{\alpha \beta \gamma \delta \epsilon}(m_1,m_2,m_3,m_4,m_5)=F_1+(F_2 +F_3) +F_4\,.  
\end{equation}
This decomposition  corresponds to an expansion of the integrand into different regions where the
integration momenta is either
of the order of the soft scale or of the hard scale. More precisely, in $F_4$ both integration momenta are
soft, whereas in $F_2,F_3$ one of these is soft and the other hard, and in $F_1$ they are both hard. The
latter terms can be absorbed by the LECs. This decomposition allows  to differentiate between the pure
IR result and the full EOMS one. At present we have performed an expansion 
of these loop functions up to order $M^5$.

For the pure IR result for $g_A$ as well as for the two-loop contributions to $Z$  we need four two-loop
functions, namely
\begin{equation}
F_{1,0,0,1,1}(M,m,M), \, F_{2,0,0,1,1}(m,M,M), \,F_{1,0,1,1,1}(M,m,m,M),  \, F_{1,1,1,1,1}(M,M,m,m,M)~,
\end{equation}
with $M$ and $m$ the leading order terms of the quark mass expansion of the pion and the nucleon mass,
respectively.
For the graphs with the axial coupling to three pions,  we need three additional loop functions
with three pion propagators instead of two:
\begin{equation}
 \,F_{1,0,1,1,1}(m,M,M,M),\,F_{1,1,1,1,1}(M,M,m,m,M),\,F_{1,1,0,0,1}(M,M,M)|_{p^2=0}~,
 \label{eq:Ffig2}
\end{equation} 
and one with two pion propagators  $F_{2,0,0,1,1}(M,m,M)$, which  is, however, not an
independent loop function as it can be expressed in terms of
$F_{1,0,0,1,1}(M,m,M)$ and $F_{2,0,0,1,1}(m,M,M)$ \cite{Kaiser:2007kf}. In the case of the second and 
third scalar integrals in Eq.~\eqref{eq:Ffig2} we only need the first term of the expansion in $M$
which we take from Ref.~\cite{Assi:2021inw}.
For the EOMS calculation more loop functions are needed which involve only one pion mass,
thus contributing only to $F_{1,2,3}$. These are 
\begin{equation}
F_{1,0,0,1,1}(M,m,m) |_{ p^2=0,m^2}, \, \, \,F_{1,0,1,1,1}(M,m,m,m)~.
\end{equation}
Indeed those necessarily  have $F_4 \equiv 0$.
One also needs the products of two one-loop functions
\begin{eqnarray}
A(m_1)&=&\int \frac{d^d k}{(2\pi)^d}  	
	\frac{1}
	{ [k^2-m_1^2]}~,
    \nonumber
    \\
    B(m_1,m_2)&=&\int \frac{d^d k}{(2\pi)^d}  	
	\frac{1}
	{ [k^2-m_1^2][(k-p)^2-m_2^2]}~,
\end{eqnarray}
which are well known in one loop calculation within BCHPT. 

Let us now turn to the contribution of one-loop graphs  with either  vertices from ${\cal L}_{\pi N}^{(3)}$ or
insertions from  mesonic operators with LECs from ${\cal L}_{\pi  \pi}^{(4)}$.  
This contribution  denoted  as $g_A^{d_i}$ is necessary to cancel 
the divergences $\sim\log( M/\mu)/\epsilon$ and $\sim\log( m/\mu)/\epsilon$ appearing in the two-loop calculation.
In fact this property was used in Ref.~\cite{Bernard:2006te}
to determine $\alpha_4$ from a renormalization group condition. Here this cancellation provides one
of the various checks of the result. 
The LECs of this part of the chiral pion-nucleon
Lagrangian are usually denoted as  $d_i$ and nine of them  contribute to 
$g_A$ at ${\cal O}(q^4)$, namely $d_{1,2,10,11,12,13,14,16,18}$. 
They  satisfy
\begin{equation}
d_{i}=\mu^{-2 \epsilon}\,\left(\delta_i \lambda+d_i(\mu)+\epsilon \, d_i^\epsilon(\mu)+ {\cal O}(\epsilon^2)\right)
\label{eq:dimu}
\end{equation}
with  the $\delta_i$   given  in 
Ref.~\cite{Gasser:2002am} for the IR case  and  in Ref.~\cite{Siemens:2016hdi} for the full EOMS
case.   Note that  each  $\delta_i$ has an expansion in powers of $g_0$,
  which up to the order we are working  has the form
\begin{equation}
\delta_i^{} = \delta_i^{(0)}+\delta_i^{(1)}  g_0+\delta_i^{(2)}  g_0^2+\delta_i^{(3)} g_0^3+\delta_i^{(4)}  g_0^4 +{\cal O}(g_0^5)\,,
\label{eq:dimug}
\end{equation}
where  the  $\delta_i$  with $i=(1,2)$/$(10-13,16)$ have non-vanishing contributions of only even/odd
powers of $g_0$ respectively, while for $i=14$ only $\delta_{14}^{(4)}$ is different from zero. This has to be kept in mind
when checking the scale dependence.
There are also contributions from the mesonic LECs $l_3$ and $l_4$,
which enter the quark mass expansion of the pion mass and the pion decay constant, respectively. Similarly to the $d_i$ one has
\begin{equation}
    l_i =\mu^{-2 \epsilon}\,\left(\gamma_i \lambda +l_i (\mu)+\epsilon\, l_i^\epsilon(\mu)+{\cal O}(\epsilon^2)\right)
    \label{eq:limu}
\end{equation}
with 
\begin{equation}
    \gamma_3=-\frac{1}{2},  \, \,   \gamma_4=2~.
\end{equation}
We have given  the LECs up to order $\epsilon$ as they will be multiplied by one-loop
terms and thus the $\epsilon$ terms can contribute at two-loop order.
The  result  of a one-loop calculation can be written in terms of  a sum of some scalar master
integrals which can be split into
an infrared part ($I$) and a regular one ($R$) with coefficients $c_{I,R}^{n}$.  One can decompose the
contributions from the insertions of the $d_i$  as
\begin{equation}
g_A^{d_i}=d_i  \sum_n (I^{n} c_I^{n} +R^{n} c_R^{n})= \sum_n I^{n} c_I^{n} d_i|_{IR}
  + \sum_n ( R^{n} c_R^{n} d_i|_{IR}+ I^{n} c_I^{n} d_i|_R )+ \sum_n R^{n}  c_R^{n} d_i|_R  \, ,
\label{eq:gadi}
\end{equation}
where $d_i|_{IR}$ contains the $\delta_i$ from Ref.~\cite{Gasser:2002am} and $d_i|_{R}$ contains
the difference between the EOMS and IR results for the $\delta_i$. The first sum in the last equality
in Eq.~\eqref{eq:gadi}  added to the IR two-loop contributions constitutes the IR result. It satisfies the
Ward identities. The same is true for the second sum together with  the mixed terms  from the two-loop graphs,
which we will call the mixed result. Finally one has the contribution from the third sum and the regular
part of the two loop graphs which can be absorbed into  $g_A^{\rm ct}$.  The EOMS result is the sum of
these three parts.

Finally, we discuss the last contribution  to the two-loop calculation of $g_A$.
In  Eq.~\eqref{eq:ga2loop}  $g_A^{\rm ct}$ denotes  counterterms generated by  a linear combination of
LECs from the Lagrangian at order $q^5$:
\begin{equation}
g_A^{\rm ct}=  \left(C^{(0)}+C^{(1)} g_0 +C^{(2)} g_0^2+C^{(3)} g_0^3+C^{(5)} g_0^5\right)\, M^4\, \equiv C\, M^4~.
\label{eq:defC}
\end{equation}
They are  necessary to absorb the remaining infinities, however, as explained before, they cannot cancel
the  $\log M/\epsilon$ terms as they are coefficients of an  analytic term in the expansion of $g_A$.
One  has
\begin{eqnarray}
C&=&\mu^{-4 \epsilon}\, \left(C_2 \lambda_2 +C_1(\mu) \lambda_1 +C_0(\mu) +{\cal O}(\epsilon)\right)~,
\nonumber \\
&=& \mu^{-4 \epsilon}\, \sum_{i=1,2,3,5}\left(C_2^{(i)}g_0^i\lambda_2 +C_1^{(i)}(\mu) g_0^i\lambda_1
+C_0^{(i)}(\mu) g_0^i +{\cal O}(\epsilon)\right)~,
\end{eqnarray}
with
\begin{eqnarray}
    \lambda_2 &=& \lambda_0^2+ \bigl(\log (4 \pi)-\gamma_E +1 \bigr)^2~,\nonumber \\
    \lambda_1 &=& \lambda_0+ \bigl(\log (4 \pi)-\gamma_E +1 \bigr)~,
\end{eqnarray}
which is appropriate for the modified $\overline{\rm MS}$ subtraction scheme used here as it is customary in CHPT
(for similar results in two-loop calculation in the  meson sector see, e.g., Ref.~\cite{Amoros:1999dp}).
Let us consider the  derivative of  $C$ with respect to $\mu$. It turns out that  the contribution to $g_0^i$
of this derivative   is not given by the derivative of $C^{(i)}$ with respect to $\mu$ as the latter can
contribute to various powers of $g_0$.  This is specific to  the baryon sector as  already encountered
in the case of the  $d_i$, see discussion after  Eq.~\eqref{eq:dimug}.
The  $C^{(i)}$  therefore can not  be  individually scale-independent. Thus  one has
\begin{equation}
\mu \frac{d}{d \mu} C =0  ~,
\end{equation}
meaning of course that each contribution  to  $g_0^i$  of the derivative of $C$  with respect to
$\mu$ is a scale-independent  quantity.
Consequently,  $C_2$ has to be scale-independent whereas the two others are scale-dependent and
satisfy the following  relations \cite{Golowich:1995kd}
\begin{equation}
\frac{d}{d \mu} C_1(\mu)= \frac{4 C_2}{\mu} ,\,\,\, \frac{d}{d \mu} C_0(\mu)=\frac{4 C_1(\mu)}{\mu}~.
\end{equation}
To get these relations provides another check of the calculation together 
with the scale-independence of the result. One  obtains
\begin{eqnarray}
C^{(0)}&=&\frac{8 \pi^2}{F^2 }\, \left(-20 d_{16}^r+8 d_{18}^r+14 d_{10}^r+8 d_{11}^r+3 d_{12}^r+d_{13}^r \right)
\,\lambda_1+ C_0^{(0)}(\mu)~,
\nonumber \\
C^{(1)}&=&-\frac{7}{ 12F^4 }\lambda_2+ \lambda_1 \, \biggl \{  -\frac{16 \pi^2 }{F^2 }(d_{14}^r-2(d_{1}^r+d_{2}^r))
-\frac{32 \pi^2}{F^4}(l_3^r-l_4^r)-\frac{323}{144 F^4}\biggr \}+ C_0^{(1)}(\mu) ~,
\nonumber \\
C^{(2)}&=&\frac{128 \pi^2}{F^2}\, \left(d_{18}^r-3 d_{16}^r\right)\,\lambda_1+ C_0^{(2)}(\mu)~,
\nonumber \\
C^{(3)}&=&\frac{7}{ 12F^4 }\lambda_2+ \biggl \{  
-\frac{64 \pi^2(l_3^r-l_4^r)}{F^4}+\frac{1}{72 F^4}(65 -32 \pi^2)\biggr \}\,\lambda_1+ C_0^{(3)}(\mu)~,
\nonumber \\
C^{(5)}&=&\frac{4}{F^4 }\lambda_2-\frac{11}{6 F^4}\lambda_1+ C_0^{(5)}(\mu)~.
\end{eqnarray}

\section{Results and discussion}
\label{sec:res}

We can now put all pieces together. We start with the  infrared result for $Z_{\rm (2-loop)}$, which reads
\begin{eqnarray}
Z_{\rm (2-loop)}|_{IR}&=&\frac{3 M^4}{ 2048 \pi^4  F^4} g_0\biggl \{ g_0^4 \biggl[ 
\frac{5 \lambda _2}{4}+\lambda _1 \left(1-5 \log\frac{M}{\mu }\right)+10 \log ^2\frac{M}{\mu }-4 \log\frac{M}{\mu }+\frac{1}{24}
   \left(6+281 \pi ^2\right)
 \biggr]\biggr.
\nonumber \\   
&&   +g_0^2 \biggl[
-3 \lambda _2+\lambda _1 \left(12 \log \frac{M}{\mu }+5\right)-24 \log ^2\frac{M}{\mu }-20 \log \frac{M}{\mu }-\frac{\pi ^2}{2}-5\biggr]
\nonumber \\   
&&   \biggl.  +\biggl[\frac{3 \lambda _2}{2}+\lambda _1 \left(2-6 \log \frac{M}{\mu }\right)+12 \log ^2\frac{M}{\mu }-8 \log \frac{M}{\mu }+\frac{1}{4}
   \left(22+\pi ^2\right) \biggr]\biggr \}~.
   \label{eq:Z2lr}
\end{eqnarray}
We give the result of the mixed terms in~\ref{app:A}.  There is in addition a contribution from the
purely regular local terms. As already stated the sum of all the regular parts satisfies the Ward
identities and thus can be absorbed in the LECs of the Lagrangian of fifth order. 
We  thus  refrain from giving any pure two-loop regular pieces here
as at the order we are working they are irrelevant. The sum of the mixed and  the pure infrared
terms (+ the regular parts)  above leads to the EOMS expression of the two-loop contribution to the $Z$ factor.
Adding the various $M^4$ contributions to $g_A$ discussed so far one gets:
\begin{equation}\label{alpha4}
\alpha_4=-\frac{7}{3} g_0\,\left(1 -g_0^2\right)+ 16 g_0^5~.
\end{equation}
This result agrees with  the $g_0^5$ term of Ref.~\cite{Bernard:2006te}. However, we found a small mistake in that 
reference. The quantity $ \tilde \alpha_4$ which takes into account the
quark mass expansion of the decay constant was too large by a factor of two, it should  read
$\tilde \alpha_4=2  \alpha_2$. Taking this into account and the fact that the result there is given in
terms of the physical pion mass our results are in agreement. 
For $\gamma_4$ we get
\begin{equation}\label{gamma41}
\gamma_4^{}= \gamma_4^{(0)}+\gamma_4^{(1)} g_0 +  \gamma_4^{(2)} g_0^2 +\gamma_4^{(3)}  g_0^3+\gamma_4^{(5)}  g_0^5~,
\end{equation}
with 
\begin{eqnarray}\label{gamma4}
 \gamma_4^{(0)}&=&16\pi^2 F^2 \,\left(-20 d_{16}^r+8 d_{18}^r+14 d_{10}^r+8 d_{11}^r+3 d_{12}^r+d_{13}^r\right)~,
 \nonumber \\
  \gamma_4^{(1)}&=& -32 \pi^2 F^2 \, \left(d_{14}^r-2(d_{1}^r+d_{2}^r)\right) -64 \pi^2 \left(l_3^r-l_4^r\right)-\frac{389}{36 }
\nonumber \\
&& \quad - \frac{64 \pi^2 F^2}{m}\biggl( c_2 +c_3- c_4
 -\frac{1}{2 m}\biggr)~, 
\nonumber \\
 \nonumber \\
  \gamma_4^{(2)}&=&256 \pi^2 F^2 \,\left( d_{18}^r-3d_{16}^r\right)~,
 \nonumber \\
 \gamma_4^{(3)}&=&-128 \pi^2\, \left(l_3^r-l_4^r\right)+\frac{1}{9}\left(13-16 \pi^2\right)+\frac{48 \pi^2 F^2}{m^2}~,
  \nonumber \\
 \gamma_4^{(5)}&=&\frac{11}{3}~.
\end{eqnarray}    
And finally one has 
\begin{equation}\label{beta4}
\beta_4^{}= \beta_4^{(0)}+\beta_4^{(1)} g_0 +  \beta_4^{(2)} g_0^2 +\beta_4^{(3)}  g_0^3+\beta_4^{(5)}  g_0^5~,
\end{equation}
with
\begin{eqnarray}\label{beta4}
(4 \pi F)^4 \beta_4^{(0)}&=&-4 \pi^2\,\left(2 d_{10} ^r+4 d_{11}^r+3 d_{12}^r+d_{13}^r\right) + C_0^{(0)}~,
\nonumber \\
(4 \pi F)^4 \beta_4^{(1)}&=&-8 \pi^2\, \left(d_{14}^r+2 (d_1^r+d_2^r)\right)-32 \pi^2 l_3^r +\frac{3575}{864}-\frac{\pi^2}{3}+\frac{1}{2}\psi^{(1)}\biggl(\frac{2}{3}\biggr) +C_0^{(1)}
\nonumber \\
&& \quad + \frac{16 \pi^2 F^2}{m}\biggl( c_2 + 4 c_4 +\frac{2}{m}\biggr)~,
\nonumber \\
(4 \pi F)^4 \beta_4^{(2)}&=&-64 \pi^2 F^2\,\left(3 d_{16}^r-d_{18}^r\right) + C_0^{(2)} \,,
\nonumber \\
(4 \pi F)^4 \beta_4^{(3)}&=&32 \pi^2\,\left(-3 l_3^r +l_4^r\right)-\frac{\pi^2}{27}\left(61 +48 \log 3\right) -\frac{335}{432}- \frac{1}{6}\psi^{(1)}\biggl(\frac{2}{3}\biggr) + C_0^{(3)}  +\frac{32 \pi^2 F^2  }{m^2}~, 
\nonumber \\
(4 \pi F)^4 \beta_4^{(5)}&=&\frac{41}{36} + \frac{7}{3} \pi^2 + C_0^{(5)}~,
\end{eqnarray}
where $\psi^{(1)}(2/3) =3.06388$ is the first derivative of the digamma function at $2/3$.
Note that the $\beta_4^{(i)}$
also have contributions from the $d_i^\epsilon$ and $l_i^\epsilon$ terms in Eqs.~\eqref{eq:dimu}
and \eqref{eq:limu}, respectively. These can be absorbed in the $C_0^{(i)}$.
In the EOMS scheme the mixed terms 
do not contribute  to $\alpha_4^{(i)}$. This has to be the case  as the leading  non-analytic terms have
to be the same in all renormalization schemes.  This is in principle  not the case for  $\gamma_4^{(i)}$
as this coefficient is renormalization scheme-dependent due to the different treatment of one-loop
diagrams in different 
renormalization schemes. It turns out that in our case the mixed terms do not contribute to  $\gamma_4^{(i)}$.
Thus the only difference between EOMS and IR at that order are terms contributing to $\beta_4$. These
will be hidden in our  $C_0$s.

\begin{figure}[t]
\begin{minipage}{0.49\linewidth}
   \includegraphics[width=1.0\linewidth]{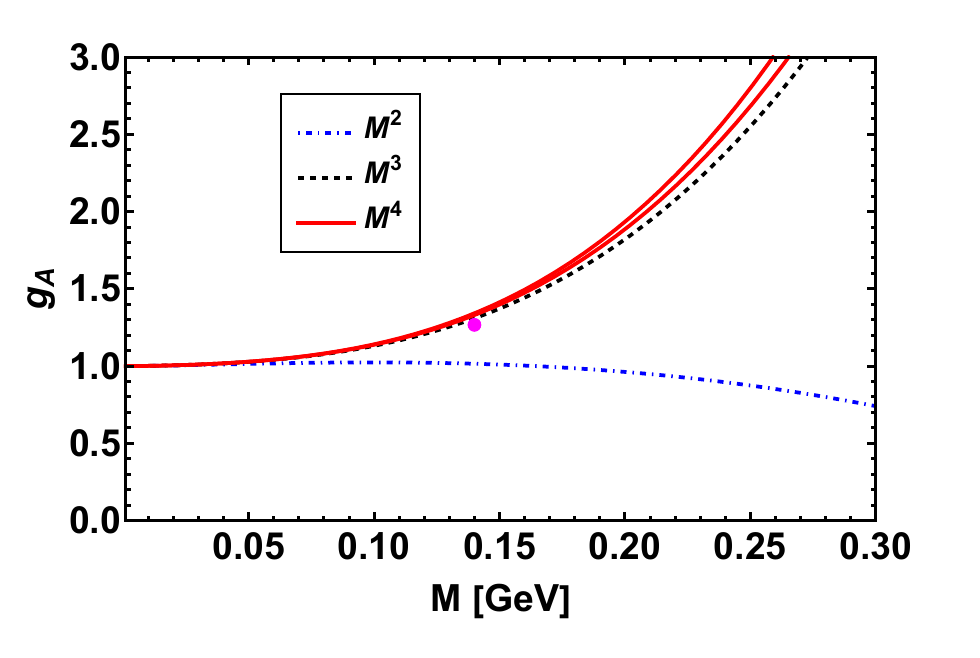}
 \end{minipage}
\hspace{0.15truecm}
\begin{minipage}{0.49\linewidth}  
     \hfill\includegraphics[width=1.0\linewidth]{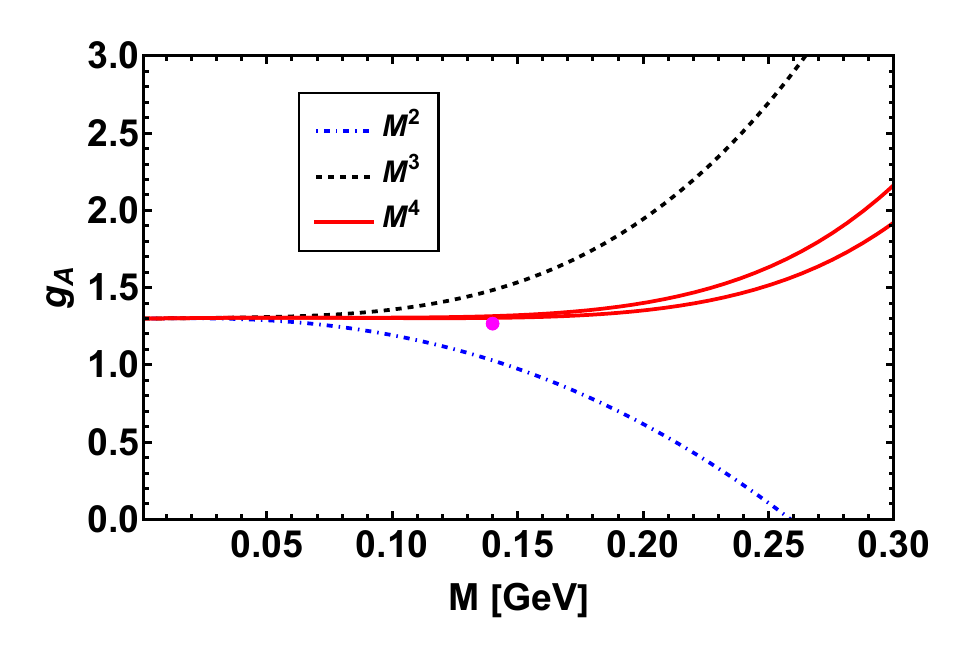}
\end{minipage}
     \vspace{-0.20cm}
     \caption{
     $g_A$ as a function of $M$ for  set~1 (left panel) and set~2 (right panel) of the  LECs
       given in the text.   The blue dash-dotted, the black dashed and the red solid lines represent the
        results up to $M^2$, $M^3$ and $M^4$, respectively. The upper $M^4$ curve corresponds to
       $C_0(m)= 15\,$GeV$^{-4}$, whereas the lower $M^4$ curve represents the case  $C_0(m)= -15\,$GeV$^{-4}$.
       The pink circle denotes the physical value of $g_A$.}
    \label{fig:ga1}
\end{figure}

Let us now discuss the  convergence of the series and the dependence of $g_A$ on the pion mass which
is relevant for the lattice. We will not perform fits to lattice data here, but rather use 
typical values for the pertinent LECs  from various analyses of elastic and inelastic pion-nucleon scattering
in the EOMS scheme to get an idea about the size of the leading two-loop corrections. A more detailed
discussion including also an uncertainty analysis will   be given in Ref.~\cite{bggmlong}.
To be concrete, we use  $F_{\pi}=0.927\,$GeV and $M_{\pi}=0.139\,$GeV,
and the nucleon mass in the chiral limit is taken as   $m=0.87\,$GeV~\cite{Hoferichter:2015hva}.
Further, we set the
renormalization scale to $\mu=m$ (which leads to $\log(m/\mu)$=0, a quantity which appears in the mixed
terms and in the regular ones in principle) and take  the values of the LECs at that scale. The
LECs in the  mesonic sector are rather well known, one has  $l_3(m)=1.4 \cdot 10^{-3}$
and $l_4(m)=3.7\cdot 10^{-3}$. From the values of $l_3$ and $l_4$ one can determine $F$ and $M$.
In the baryon sector  the  LECs are less known. From an analysis of elastic and inelastic
pion-nucleon scattering~\cite{Siemens:2017opr} 
we take set~1 (which is based on the standard power counting $m_N\sim \Lambda_\chi$)
$c_2=3.51\,$GeV$^{-1}$, $c_3=-6.63\,$GeV$^{-1}$,  $c_4=4.01\,$GeV$^{-1}$,
$\bar d_1+\bar d_2=4.37\,$GeV$^{-2}$, 
$\bar{d}_{10}= -0.8\,$GeV$^{-2}$,
$\bar{d}_{11}= -15.6\,$GeV$^{-2}$,
$\bar{d}_{12}= 5.9\,$GeV$^{-2}$,
$\bar{d}_{13}= 13.6\,$GeV$^{-2}$,
$\bar{d}_{14}= -7.43\,$GeV$^{-2}$,
and
$\bar{d}_{16}= 0.4\,$GeV$^{-2}$,
and as set~2 (which is based on the power counting used in nucleon-nucleon scattering
$m_N\sim  \Lambda_\chi^2/M_\pi$),
$c_2=4.89\,$GeV$^{-1}$, $c_3=-7.26\,$GeV$^{-1}$,  $c_4=4.74$ GeV$^{-1}$,
$\bar d_1+\bar d_2=3.39,$GeV$^{-2}$, 
$\bar{d}_{10}= 10.9\,$GeV$^{-2}$,
$\bar{d}_{11}= -30.9\,$GeV$^{-2}$,
$\bar{d}_{12}= -10.9\,$GeV$^{-2}$,
$\bar{d}_{13}= 27.7,$GeV$^{-2}$,
$\bar{d}_{14}= -7.36\,$GeV$^{-2}$,
and
$\bar{d}_{16}= -3.0\,$GeV$^{-2}$.
Note that we only have information on  $\bar{d}_{14} -\bar d_{15}$  and we have assumed here $\bar d_{15}=0$.
Further, the $\bar{d}_i$ are the $d_i^r (\mu)$ defined at $\mu = M_\pi$, see, e.g., Ref.~\cite{Fettes:1998ud}.
These central values have been obtained in heavy baryon chiral perturbation theory from a combined fit
to the reactions $\pi N \to \pi N$ and $\pi N \to \pi \pi N$.  As we have done an expansion in $M/m$ of
the IR and  EOMS expressions
our results should in fact correspond to the one in the heavy baryon approach.
Finally,  $\bar{d}_{18}= -0.8\,$GeV$^{-2}$ can be related to the Goldberger-Treiman discrepancy
\cite{Fettes:1998ud,Becher:2001hv}. For the LEC $C_0\equiv \sum_i C_0^{(i)} g_0^i$ we assume it to be of natural size,
typically of the order of $1/\Lambda^4_\chi$ with $\Lambda_\chi=0.6\,$GeV an estimate of the
breakdown scale of the chiral expansion (which is a more conservative estimate than given above and was
used in Refs.~\cite{Siemens:2016hdi,Siemens:2017opr}).
In Fig.~\ref{fig:ga1} (left panel) we show the chiral expansion  of $g_A$ at second,
third and fourth order for set~1 of the LECs
and similarly for set~2 in the right panel of Fig.~\ref{fig:ga1}.
The upper fourth order curve corresponds to  $ C_0(m)= 15\,$GeV$^{-4}$, whereas the lower
fourth order curve represents the case  $C_0(m)=-15\,$GeV$^{-4}$.
While the third order correction  is large, as the coefficient $\alpha_3$ in Eq.~\eqref{gAstruc}
is proportional to the large factor $2c_4-c_3$, we see that the fourth
order corrections are rather small for pion masses below 300~MeV, even though some of   the
dimension-three LECs are rather large.  Also, we find that $g_0 = 1.0$ for  set~1 and $g_0=1.3$
for set~2, in order.
\noindent Furthermore, at the physical  pion mass one has for $C_0^{(i)}(m)= 0$ 
 \begin{eqnarray}
{\rm set}~1 &:& \Delta^{(2)}=  1.5 \%~, \,~~~\quad  \Delta^{(3)} =28.8 \%~, \quad  \Delta^{(4)}=2.6\%~,\nonumber\\
{\rm set}~2 &:& \Delta^{(2)}=  -26.7 \%~, \quad  \Delta^{(3)} =44.5 \%~, \quad  \Delta^{(4)}=-17.4\%~, 
\end{eqnarray}
which shows the same pattern as discussed above, namely large corrections at  order  $q^3$ for
both sets of the LECs, but small/moderate ones at leading two-loop order $q^4$ for set~1 and set~2, respectively. 
We note that the quantity $\Delta^{(2)}$ is rather sensitive to the actual value of $\bar{d}_{16}$.
Furthermore, we point out that in the recent
lattice QCD fits in Ref.~\cite{Hall:2025ytt}, $g_0$ comes out in the range $1.26-1.30$.

\section{Summary and outlook}
\label{sec:summ}

In this paper, we have calculated and analyzed the leading two-loop corrections to the nucleon
axial-vector coupling $g_A$, that is of fundamental importance in low-energy QCD. We used  two
covariant versions  of baryon chiral perturbation theory, namely we applied the EOMS and the IR renormalization
schemes. The pertinent expression for the fourth order corrections $\sim M^4$, denoted as $\alpha_4$,
$\gamma_4$ and $\beta_4$, as defined in Eq.~\eqref{gAstruc}, are explicitly given 
in Eq.~\eqref{alpha4}, Eq.~\eqref{gamma4} and Eq.~\eqref{beta4}, respectively. We have used
two sets of  the dimension-two
and dimension-three LECs from a combined study of the $\pi N \to \pi N$ and  $\pi N \to \pi\pi N$ processes in the
EOMS scheme and calculated the fourth order corrections as shown in Fig.~\ref{fig:ga1}. For set~1 of the
LECs,  the corrections turned out to be rather small, signaling a good convergence of the chiral 
expansion of $g_A$, whereas for set~2 they are larger but still moderate.
However, to finally draw conclusions, the  $M^5$ corrections
 need to be worked out and a more detailed uncertainty analysis has to 
be performed. Such work is in progress.

\section*{Acknowledgements}
This work was supported in part by the European
Research Council (ERC) under the European Union's Horizon 2020 research
and innovation programme (grant agreement No. 101018170).
The work of UGM was also supported in part by the CAS President's International
Fellowship Initiative (PIFI) (Grant No. 2025PD0022), by the MKW NRW
under the funding code NW21-024-A and by by the Deutsche Forschungsgemeinschaft (DFG,
German Research Foundation) as part of the CRC 1639 NuMeriQS – project
no. 511713970.
JG acknowledges support by the Georgian Shota Rustaveli National Science Foundation (Grant No. FR-23-856).

\begin{appendix}
\section{Mixed terms in the wave function renormalization factor at two loops }
\label{app:A}
We give the expression of $Z_{(\rm 2-loop)}|_{\rm mix}$, the contribution from the mixed terms $F_2 +F_3$ of the loop
functions,
 \begin{eqnarray}
Z_{(\rm 2-loop)}|_{\rm mix}&=&\frac{3 M^4}{32 \pi^4 F^4} g_0\biggl \{ g_0^4 \biggl(  \frac{9 \lambda _2}{32}+\lambda _1 \left(-\frac{9}{16} \log \frac{m}{\mu }-\frac{9}{16} \log \frac{M}{\mu } +\frac{71}{192}\right)+\log
   \frac{m}{\mu } \left(\frac{9}{8} \log \frac{M}{\mu }-\frac{71}{96}\right)\biggr.
   \nonumber \\
 &&  \quad  +\frac{9}{16} \log ^2\frac{m}{\mu }+\frac{9}{16} \log
   ^2\frac{M}{\mu }-\frac{71}{96} \log \frac{M}{\mu }+\frac{1}{384} \left(113+18 \pi ^2\right)
\nonumber \\   
&&   +g_0^2 \biggl(
-\frac{5 \lambda _2}{64}+\lambda _1 \left(\frac{5}{32} \log \frac{m}{\mu }+\frac{5}{32} \log \frac{M}{\mu }-\frac{37}{64}\right)+\log
   \frac{m}{\mu } \left(\frac{37}{32}-\frac{5}{16} \log \frac{M}{\mu }\right)
   \nonumber \\
&&  \quad -\frac{5}{32} \log ^2\frac{m}{\mu }-\frac{5}{32} \log
   ^2\frac{M}{\mu }+\frac{37}{32} \log \frac{M}{\mu }-\frac{5}{384} \left(36+\pi ^2\right)
\biggr)
\nonumber \\   
&&    +\biggl(
-\frac{3 \lambda _2}{64}+\lambda _1 \left(\frac{3}{32} \log \frac{m}{\mu }+\frac{3}{32} \log \frac{M}{\mu }-\frac{5}{128}\right)+\log
   \frac{m}{\mu } \left(\frac{5}{64}-\frac{3}{16} \log \frac{M}{\mu }\right)
   \nonumber \\
&& \quad \biggl. -\frac{3}{32} \log ^2\frac{m}{\mu }-\frac{3}{32} \log
   ^2\frac{M}{\mu }+\frac{5}{64} \log \frac{M}{\mu }+\frac{1}{256} \left(-33-2 \pi ^2\right)
\biggr \}~.
\end{eqnarray}
One has, see remark after Eq.~\eqref{eq:Z2lr},
\begin{equation}
Z_{(\rm 2-loop)}|_{\rm EOMS}=Z_{(\rm 2-loop)}|_{\rm IR}+Z_{(\rm 2-loop)}|_{\rm mix} +(Z_{(\rm 2-loop)}|_{\rm reg})~,
\end{equation}
where we put the regular piece in brackets as it is anyway absorbed by the LECs. 

\end{appendix}


\vfill
\eject
%
\vspace{1.5truecm}

\vspace{1.5truecm}

\end{document}